\documentclass[12pt,a4paper]{article}%

\usepackage{cite}
\usepackage{amsmath,amssymb,amsfonts}
\usepackage{calc}
\usepackage{hyperref}
\usepackage[pdftex]{color,graphicx}

\numberwithin{equation}{section} 
\newcommand{\hhref}[1]{\href{http://arxiv.org/abs/#1}{arXiv:#1}}
\newcommand{\beq}{\begin{equation}}
\newcommand{\eeq}{\end{equation}}

\begin{document}
\hfill CERN-PH-TH/2012-086
\renewcommand{\thefootnote}{\fnsymbol{footnote}}
\color{black}
\vspace{0.5cm}
\begin{center}
{\Huge \bf (Dys)$Z$philia}\\[5mm] {\LARGE \bf or} \\[5mm] { \LARGE \bf a custodial breaking Higgs at the LHC} \\
\bigskip\color{black}\vspace{0.6cm}
{{\large\bf Marco Farina$^{a,b}$,    Christophe Grojean$^{a}$ and  Ennio Salvioni$^{a,c,}$\footnote{Email: \url{marco.farina@sns.it},~ \url{Christophe.Grojean@cern.ch},~ \url{Ennio.Salvioni@cern.ch}}}
} \\[7mm]
{\it  (a)  Theory Division, Physics Department, CERN, CH-1211 Geneva 23, Switzerland}\\[3mm]
{\it  (b)  Scuola Normale Superiore and INFN, Piazza dei Cavalieri 7, I-56126 Pisa, Italy}\\[3mm]
{\it  (c)  Dipartimento di Fisica and INFN, Universit\`a di Padova, Via Marzolo 8,\\[1.5mm] I-35131 Padova, Italy}\\[3mm]
\end{center}
\bigskip
\centerline{\large\bf Abstract}
\begin{quote}
Electroweak precision measurements established that custodial symmetry is preserved to a good accuracy in the gauge sector after electroweak symmetry breaking. However, recent LHC results might be interpreted as pointing towards Higgs couplings that do not respect such symmetry. Motivated by this possibility, we reconsider the presence of an explicitly custodial breaking coupling in a generic Higgs parameterization. After briefly commenting on the large UV sensitivity of the $T$ parameter to such a coupling, we perform a fit to results of Higgs searches at LHC and Tevatron, and find that the apparent enhancement of the $ZZ$ channel with respect to $WW$ can be accommodated. Two degenerate best-fit points are present, which we label `$Z$philic' and `dys$Z$philic' depending on the sign of the $hZZ$ coupling. Finally we highlight some measurements at future linear colliders that may remove such degeneracy.

\end{quote}

\renewcommand{\thefootnote}{\arabic{footnote}}\setcounter{footnote}{0}

\pagestyle{empty}

\newpage

\section{Introduction }
\setcounter{page}{1}
\pagestyle{plain}

The main goal of the LHC is to shed light on the mechanism of ElectroWeak Symmetry Breaking (EWSB). The recent excesses observed in searches for the Higgs boson at ATLAS and CMS, supplemented by some hints from the Tevatron, can be seen as the starting point in this direction. Even though they are far from being conclusive, the experimental results point to a resonance with mass around $125\,\mathrm{GeV}$ and, broadly speaking, Higgs-like behavior. If such hints really correspond to the first manifestation of a new degree of freedom, then the measurement and study of its properties will be crucial to unveil EWSB. This is even more true in the absence of any direct evidence of physics beyond the Standard Model (SM) so far.
\newline

The EWSB sector has been indirectly probed by the LEP precision tests, which represent a primary source of information: one of the most important outcomes of precision measurements is that the gauge sector after EWSB must approximately respect an $SU(2)_c$ custodial symmetry. Such requirement is satisfied by the SM description of EWSB. On the other hand, at the moment experimental excesses at the LHC may be interpreted as pointing to non-SM Higgs couplings, especially in the gauge sector. In fact, not only is there a trend of underproduction in the $WW$ channel and of overproduction in the $\gamma \gamma$ channel (for the latter, the excess is stronger in the vector boson fusion subchannel), but an enhancement of the $ZZ$ signal with respect to $WW\,$ is observed, whereas  custodial symmetry implies that the two have the same strength (when normalized to their SM values).
\newline

Clearly such hints could be just due to statistical fluctuations, or to issues with the modeling of complex backgrounds (for example, in the $h\to WW$ channel). Nevertheless, it is interesting to ask what would be the implications if the current pattern of excesses were to be confirmed with more data. In this spirit, we relax the assumption of custodial invariance in the couplings of the Higgs resonance and perform a fit to the results of Higgs searches by employing a parameterization where explicit custodial breaking is allowed. Our model-independent approach is similar in spirit to other recent analyses of the Higgs experimental results, see Refs.~\cite{Carmi:2012yp,Azatov:2012bz,Espinosa:2012ir,Giardino:2012ww,Ellis:2012rx} \footnote{See Refs.~\cite{Zeppenfeld:2000td,Plehn:2001nj,Duhrssen:2004cv,Duehrssen2003,Burgess:1999ha,Lafaye:2009vr,Bock:2010nz} for earlier studies on the determination of Higgs couplings, and Refs.~\cite{Cervero:2012cx,Li:2012ku,Rauch:2012wa} for other recent related work.}. We also analyze the effects on the electroweak parameter $T$, finding that if the couplings $hWW$ and $hZZ$ do not respect custodial symmetry, then $T$ receives quadratically divergent corrections. In a concrete model, new degrees of freedom below the cutoff must therefore conspire to make the total contribution to $T$ compatible with electroweak precision tests (EWPT).

Not surprisingly, the fit to the results of Higgs searches points to a Higgs coupling more strongly to $ZZ$ than to $WW$. Two exactly degenerate best-fit points appear, which we label `$Z$philic' and `dys$Z$philic' depending on the sign of the $hZZ$ coupling. Such sign, although unobservable in current Higgs searches, is physical in processes involving interference. We therefore discuss some future measurements at colliders that may be used to resolve the degeneracy.
\newline

We remark that many proposals for physics beyond the SM exist in the literature where the custodial symmetry is not respected: for example, models where the Higgs sector is extended with scalar triplets that get a non-vanishing vacuum expectation value, generic two Higgs doublet models, as well as theories where the Higgs arises as the pseudo-Goldstone boson of a coset $G/H$ where $H$ does not contain $SO(4)\sim SU(2)\times SU(2)$, such as $SU(3)/(SU(2)\times U(1))$, fall in this class.
\newline

Our paper is structured as follows: we start by introducing our parameterization and discussing the fit to LHC data in Section 2, where we also briefly comment on the effect of explicit custodial breaking on the electroweak $T$ parameter. In the light of our results, we discuss in Section 3 some implications for future precision measurements of Higgs properties. Finally, we conclude in Section 4.
\section{Lagrangian, $T$ parameter and fit to LHC data} \label{sec:potential_allowed_region}
We employ the usual parameterization of interactions of SM fields with a generic Higgs boson by considering an EW chiral Lagrangian coupled to a scalar resonance $h$.
The Goldstone bosons corresponding to the longitudinal polarizations of the $W$ and $Z$ are introduced through the chiral field
\begin{equation}
\Sigma(x) = \exp(i\sigma^a \pi^a(x)/v)
\end{equation}
with $v \simeq 246$ GeV. The Lagrangian mass terms are then
\begin{equation}
\label{eq:Lmass}
 {\cal L}_{mass} = \frac{v^2}{4} \text{Tr}\left[ \left( D_\mu \Sigma \right)^\dagger
 \left( D^\mu \Sigma \right) \right] - \frac{v}{\sqrt{2}} \sum_{i,j} \left( \bar u_L^{(i)} d_L^{(i)} \right)
 \Sigma \begin{pmatrix} \lambda_{ij}^u\,  u_R^{(j)} \\[0.1cm] \lambda_{ij}^d\,  d_R^{(j)} \end{pmatrix} + \mathrm{h.c.}
\end{equation}
where
\begin{equation}
D_\mu \Sigma = \partial_\mu \Sigma
 -i g \frac{\sigma^a}{2} W^a_\mu \Sigma + i g^\prime \Sigma \frac{\sigma^3}{2} B_\mu \, .
\end{equation}
We omit for simplicity lepton masses, which could be introduced in the same way as for quarks. Notice that this Lagrangian is approximately invariant under a global $SU(2)_L\times SU(2)_R$, under which $\Sigma$ transforms as
%
\begin{equation}
\Sigma \to U_L \, \Sigma\, U_R^\dagger \, .
\end{equation}
This invariance is broken in the vacuum to the diagonal $SU(2)_{c}$ (the `custodial symmetry'), which guarantees that the $\rho$ parameter, defined as
\begin{equation}
\rho = \frac{m_{W}^{2}}{m_{Z}^{2}\cos^{2}\theta_{W}} = 1+\hat{T} = 1 + \alpha T\,,
\end{equation}
satisfies the tree level relation $\rho=1$, as experimentally verified to good accuracy. In principle the Lagrangian (\ref{eq:Lmass}) could contain an additional term
%
\begin{equation}
\label{eq:CuBreak}
 v^2 \,\left( \text{Tr}\left[ \Sigma^\dagger D_\mu \Sigma \, \sigma^3 \right]\right)^2
\end{equation}
that is gauge invariant, but explicitly breaks $SU(2)_{L}\times SU(2)_{R}$ and therefore the custodial symmetry. To prevent large deviations from $\rho=1$ and thus tensions with precision tests, its coefficient has to be very small, $O(10^{-3})$, so the term (\ref{eq:CuBreak}) is usually neglected.

\par
As it is well known, the description \eqref{eq:Lmass} leads to amplitudes for longitudinal gauge boson scattering that grow with energy, and as a consequence to a loss of perturbative unitarity at a scale \mbox{$4\pi v\sim 3\,\mathrm{TeV}\,$}. To moderate the growth of amplitudes and therefore postpone the perturbative unitarity breakdown, a scalar resonance transforming as a singlet under the custodial symmetry can be introduced. We can thus add to Eq.~(\ref{eq:Lmass}) all possible interactions with the scalar resonance up to second order, obtaining \cite{Contino:2010mh} (see also Refs.~\cite{ContinoTASI,ChristopheHCP} for an introduction)
\begin{equation}
\label{eq:Lag}
\begin{split}
{\cal L}_{h} =& \frac{1}{2} \left(\partial_\mu h\right)^2 - V(h) +
 \frac{v^2}{4} \text{Tr}\left[ \left( D_\mu \Sigma \right)^\dagger \left( D^\mu \Sigma \right) \right]
 \left( 1 + 2 a\, \frac{h}{v} + b\, \frac{h^2}{v^2} + \dots \right) \\[0.1cm]
 & - \frac{v}{\sqrt{2}} \sum_{i,j} \left( \bar u_L^{i} d_L^{i} \right)  \Sigma \left( 1+ c\, \frac{h}{v} + c_{2}\,\frac{h^{2}}{v^{2}}+\cdots\right)
 \begin{pmatrix} \lambda_{ij}^u\,  u_R^{j} \\[0.1cm] \lambda_{ij}^d\,  d_R^{j} \end{pmatrix} + \mathrm{h.c.}
\end{split}
\end{equation}
where $a,b,c,c_{2}$ are free parameters (the SM is retrieved by choosing $a=b=c=1\,$, $c_{2}=0$ and vanishing terms of higher order in $h$). We do not write explicitly the scalar self-interactions contained in $V(h)$, as they will not be relevant in our discussion.

Since we are interested in custodial breaking effects, we add to the Lagrangian the following terms
\begin{equation}
\label{eq:Lcb}
{\cal L}_{cb} = -\frac{v^2}{8} \, \left(\text{Tr}\left[ \Sigma^\dagger D_\mu \Sigma \, \sigma^3 \right]\right)^2 \left(t_{cb}+2 a_{cb} \frac{h}{v}+\cdots\right),
\end{equation}
where $t_{cb}$ and $a_{cb}$ are free parameters\footnote{Higher orders in the Higgs are negligible for our purposes.} and the overall normalization has been chosen for later convenience. As we already mentioned, $t_{cb}$ contributes to $T$ at tree level, $\hat{T}=-t_{cb}\,$. On the other hand, the consequences of the coupling $a_{cb}$ can be seen by going to the unitary gauge, \mbox{$\Sigma=1$}: the interactions of the Higgs with vector bosons are modified as follows
\begin{equation}
\label{eq:Lhvv}
{\cal L}_{hVV} = \left[a \mbox{ } m_W^2 W_\mu^+ W_\mu^- +\frac{1}{2} (a+a_{cb}) \mbox{ } m_Z^2 Z_\mu Z_\mu \right]  \left(2 \frac{h}{v}\right).
\end{equation}
Clearly the ratio between the two couplings differs from the usual SM value \mbox{$g_{hWW}/g_{hZZ}=\cos^{2}\theta_W$}. In a SILH Lagrangian \cite{SILH}, where the SM gauge symmetries are linearly realized in the strong sector, we can consider the following operators
\begin{equation}
\mathcal{O}_{H}=\frac{c_{H}}{2f^{2}}\partial^{\mu}(H^{\dagger}H)\partial_{\mu}(H^{\dagger}H)\,,\qquad \mathcal{O}_{T}=\frac{c_{T}}{2f^{2}}\left(H^{\dagger}D_{\mu}H-(D_{\mu}H)^{\dagger}H\right)^{2}
\end{equation}
where $H$ is the (composite) Higgs doublet emerging as a pseudo-Goldstone boson from the strong sector. We find
\begin{equation} \label{silh}
a=1-\frac{c_{H}}{2}\frac{v^{2}}{f^{2}}\,,\qquad a_{cb}=-2c_{T}\frac{v^{2}}{f^{2}}\,.
\end{equation}
However, in addition a contribution $t_{cb}=-c_{T}(v^{2}/f^{2})$ is generated, or equivalently a correction $\hat{T}=c_{T}(v^{2}/f^{2})\,$. Therefore in this case the coefficients $t_{cb}$ and $a_{cb}$ in Eq.~\eqref{eq:Lcb} are of the same order. We recall that $c_{H}$ is in general\footnote{The contribution to $c_{H}$ arising from integrating out triplet scalars is negative. However, in models where the collective symmetry breaking mechanism is realized, such as Little Higgs theories, the total contribution to $c_{H}$ is positive even in presence of scalar triplets. See Ref.~\cite{LowRattazziVichi}.} positive definite \cite{LowRattazziVichi}, implying the generic expectation $a<1$ in composite Higgs models. However, in the following we will not restrict ourselves to this range. For a discussion of how $a>1$ could arise, see Ref.~\cite{Falkowski:2012vh}.


\subsection{$T$ parameter} \label{Tparameter}
\begin{figure}[t]
 \begin{center}
   \includegraphics[width=.6\textwidth]{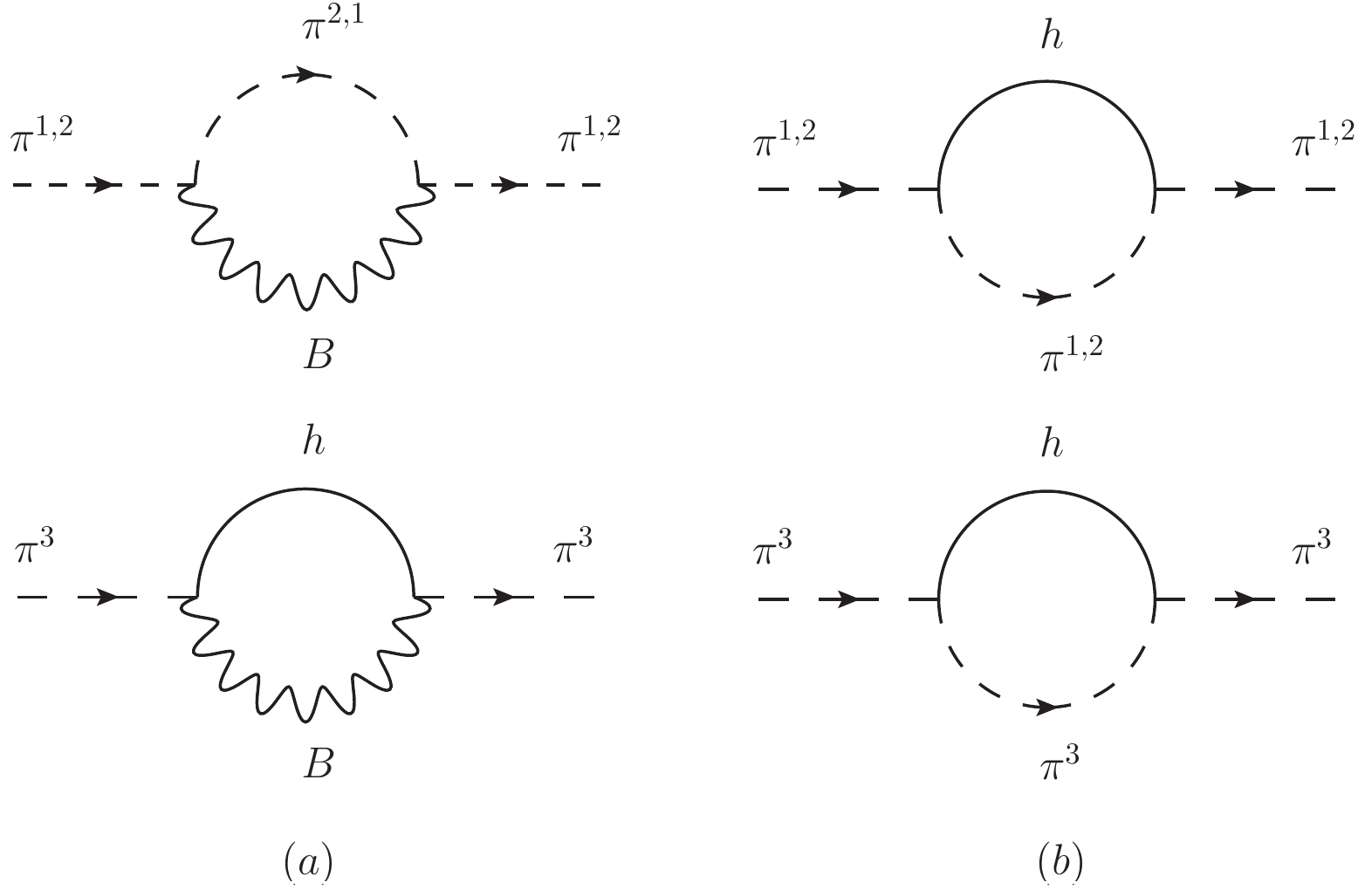}
 \end{center}
\caption{$(a)$ Diagrams giving a logarithmic divergence in $T$ when $a\neq 1\,$. This is the leading correction in the custodial-preserving case. $(b)$ Diagrams giving a quadratic divergence in $T$ when $a_{cb}\neq 1\,$, see Eq.~\eqref{quad_div}.}
\label{fig:Feynmandiagrams}
\end{figure}

\begin{figure}[h]
 \begin{center}
   \includegraphics[width=.45\textwidth]{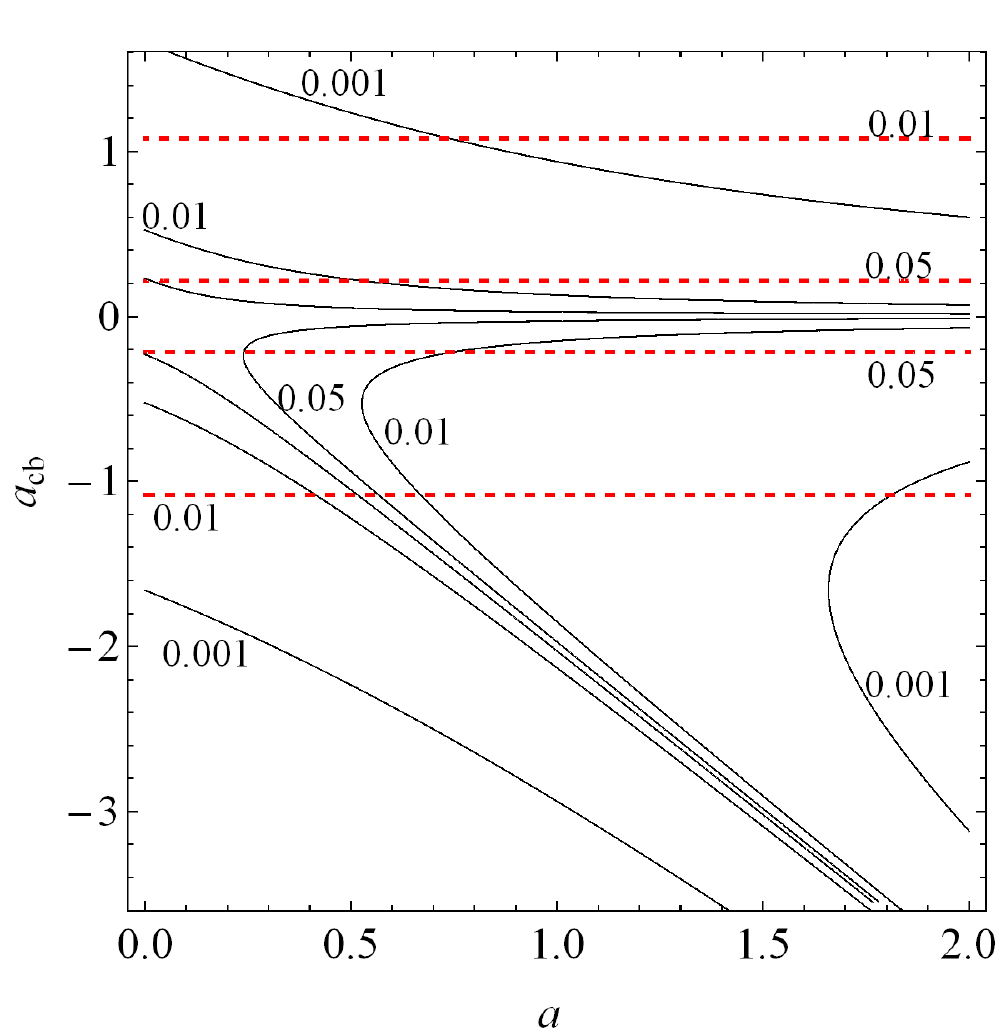}
 \end{center}
\caption{Isocontours in the $(a,a_{cb})$ plane of $|\Delta\epsilon_{1}^{UV}/\epsilon_{1}^{exp}|^{-1}\,$ (solid, black) and of $|\Delta\epsilon_{1}^{TL}/\epsilon_{1}^{exp}|^{-1}\,$ (red, dashed), roughly representing the amount of tuning needed to satisfy EWPT.}
\label{fig:resultsEWPT}
\end{figure}

It is well known that when $a\neq 1$ in Eq.~\eqref{eq:Lag}, a logarithmically divergent contribution to $T$ (as well as to $S$) arises. Such contribution is due to the diagrams in Fig.~\ref{fig:Feynmandiagrams}$(a)$, and is computable within the low-energy theory, see Ref.~\cite{BarbieriETAL}. However, in the present case we also need to consider the effects of explicit custodial breaking contained in Eq.~\eqref{eq:Lcb}. Even if we set $t_{cb}=0$, a quadratic UV sensitivity appears in $T$, due to the diagrams involving the Higgs shown in Fig.~\ref{fig:Feynmandiagrams}$(b)$. This quadratic divergence reads
\begin{equation} \label{quad_div}
\hat{T}^{UV}=\Delta\epsilon_{1}^{UV}=\frac{1}{16\pi^{2}}\frac{\Lambda^{2}}{v^{2}}\left(a^{2}-(a+a_{cb})^{2}\right)\,,
\end{equation}
where $\Lambda$ is the cutoff: setting $\Lambda = 4\pi v$, we obtain a contribution of tree-level size. In a concrete model, new degrees of freedom below the cutoff will need to conspire to make the total contribution to $T$ compatible with EW precision data. This will require in general a certain amount of tuning, which we quantify in Fig.~\ref{fig:resultsEWPT} by showing isocontours of $|\Delta\epsilon_{1}^{UV}/\epsilon_{1}^{exp}|^{-1}\,$, where the experimental value of the $\epsilon_{1}$ parameter is $\epsilon_{1}^{exp} = (5.4\pm 1.0)\times 10^{-3}\,$ \cite{lepEWWG}. In the same figure we also show isocontours of $|\Delta\epsilon_{1}^{TL}/\epsilon_{1}^{exp}|^{-1}\,$, where
\begin{equation}
\Delta\epsilon_{1}^{TL}=-\frac{a_{cb}}{2}\,
\end{equation}
is the tree-level contribution that arises when the full gauge invariant operator $\mathcal{O}_{T}$ is considered. We see that the level of tuning is roughly similar in the two cases. A full computation of $T$ requires choosing a complete model, see Refs.~\cite{Chankowski:2006jk,Chen:2006pb,Chankowski:2006hs} and references therein for examples.
%

\subsection{Recent LHC results}
In this section we will perform a fit to the results of experimental searches for the Higgs at LHC and at Tevatron. We are going to use the full set of data released in March by ATLAS \cite{ATLAS2012,ATLAS2012FP}, CMS \cite{CMS2012,CMSgg2012} and Tevatron \cite{TEVNPH:2012ab}, as reported in Fig.~3 \footnote{We have included all the channels for which a signal strength has been provided. We exclude from this set only the ATLAS results on $h \to b\bar{b}$ and $h \to \tau \bar{\tau}\,$, which are difficult to interpret within the framework of our simple analysis.}. Experimental results are given in terms of
\beq
\mu= \frac{(\sigma_{prod} \times BR)^{obs}}{(\sigma_{prod} \times BR)^{SM}} .
\eeq
In presence of a signal a best fit for this quantity is given along with errors. Several comments are in order about the dependence of $\mu$ on $(a,a_{cb},c)\,$ for the different channels:
\begin{itemize}
  \item The $p p \to h jj \rightarrow \gamma \gamma jj$ sample at CMS is assumed to be produced through Vector Boson Fusion (VBF) with a small contamination coming from gluon fusion \cite{Chatrchyan:2012tw}, so that
  \beq
  r_{\gamma\gamma jj}(a, a_{cb},c)=\frac{\sigma_{prod}(a, a_{cb},c)}{\sigma_{prod}^{SM}}= \frac{0.033 \, c^2 \, \sigma_{gg}+r_{VBF}(a,a_{cb})\, \sigma_{VBF}}{0.033 \, \sigma_{gg}+\sigma_{VBF}}
  \eeq
where
\begin{equation}
r_{VBF}(a,a_{cb})=\frac{a^{2}+R_{VBF}(a+a_{cb})^{2}}{1+R_{VBF}}\,.
\end{equation}
$R_{VBF}\sim1/2.93$ is the ratio between $ZZ$ and $WW$ fusion production in the SM (at LHC, 7 TeV) \cite{VBFatNNLO}, $\sigma_{gg}$ is the gluon fusion production cross section and \mbox{$\sigma_{VBF}/\sigma_{gg} \approx 0.079$}.

  \item  We include the ATLAS results from fermiophobic (FP) Higgs searches in \mbox{$p p \to h X \rightarrow \gamma \gamma X$}. Following Ref.~\cite{Giardino:2012ww} we take the production to be dominated by VBF with a sizable contamination from gluon fusion:
  \beq
  r_{FP}(a, a_{cb},c)= \frac{0.3 \, c^2 \, \sigma_{gg}+r_{VBF}(a,a_{cb})\, \sigma_{VBF}}{0.3 \, \sigma_{gg}+\sigma_{VBF}}\,.
  \eeq

  \item $h \rightarrow b \bar{b}$ is observed in the associated production channel $V \rightarrow V h \rightarrow V b \bar{b}$. Taking into account the possibility of $V$ being either a $W$ or a $Z$ we have
       \beq
  r_{Vh}(a, a_{cb})= \frac{a^2 + R_{Vh} (a+a_{cb})^2}{1+R_{Vh}}
  \eeq
  where $R_{Vh}$ is the ratio of $Zh$ to $Wh$ production in the SM, equal to $0.55$ at LHC and to $0.61$ at Tevatron.

  \item All the other channels are assumed to come from inclusive production. In this case for LHC
  \beq \label{LHCinclprod}
  r_{incl}^{LHC}(a, a_{cb},c)= \frac{c^2 \sigma_{gg}+r_{VBF}(a,a_{cb})\, \sigma_{VBF}+r_{Vh}(a,a_{cb})\,\sigma_{Vh}}{ \sigma_{gg}+ \sigma_{VBF}+\sigma_{Vh}} \sim c^2
  \eeq
  where $\sigma_{Vh}/\sigma_{gg}\approx 0.058$, and the last approximate equality holds because the main production mechanism is gluon fusion. We have checked that considering inclusive $WW$ and $ZZ$ production as coming only from gluon fusion and VBF, as done in Ref.~\cite{Espinosa:2012ir}, does not significantly affect our results. An equation completely analogous to \eqref{LHCinclprod} holds for inclusive production at Tevatron.

  \item The partial width for $h\to \gamma\gamma$, which arises both from $W$ and from heavy fermion (top, bottom and tau) loops, gets rescaled as
  \begin{equation}
  r_{\gamma\gamma}(a)=\frac{\Gamma(h\to \gamma\gamma)}{\Gamma(h\to \gamma\gamma)_{SM}}\simeq (1.26\,a - 0.26\, c)^{2}
  \end{equation}
  for $m_{h}=125\,\mathrm{GeV}\,$.
\end{itemize}
\begin{figure}[t]
\hspace{0.3in}
\begin{minipage}[c]{0.4\linewidth}
   \includegraphics[scale=0.8]{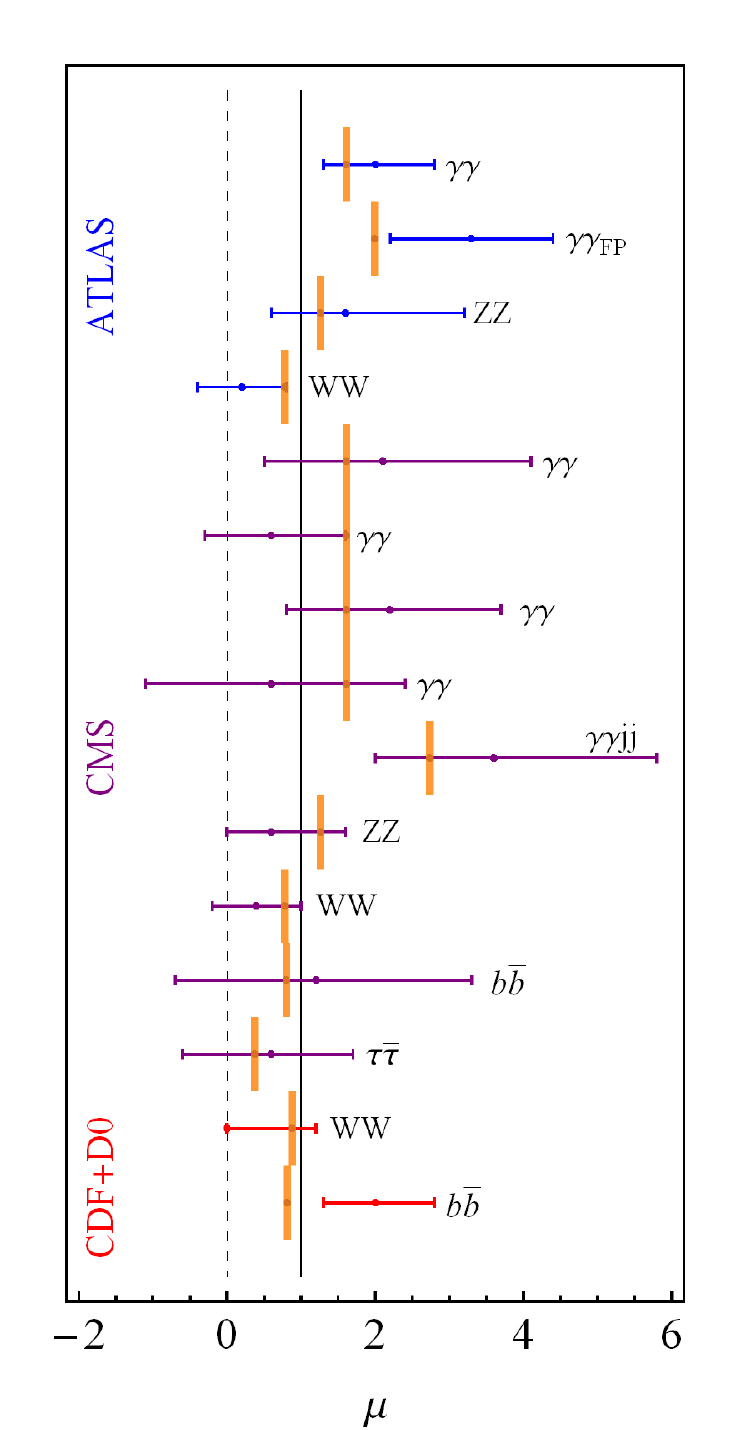}
\end{minipage}
\begin{minipage}[c]{0.5\linewidth}
\begin{tabular}{c|c|l}
\setlength{\tabcolsep}{5pt}
Channel   & $\hat{\mu} $  & $ \sim \sigma_{prod} \times  \Gamma$
\\
\hline
$ $ &$ $ &$ $
\\[-10pt]
$ \gamma \, \gamma \,\,$  & $2^{+0.8}_{-0.7} $ & $ \phantom{x} c^2 \times r_{\gamma\gamma}(a,c)$
\\[2pt]
$ \gamma \, \gamma_{FP} \,\,$  & $3.3^{+1.1}_{-1.1} $ & $ \phantom{x} r_{FP}(a, a_{cb},c) \times r_{\gamma\gamma}(a,c)$
\\[2pt]
$ Z \, Z^\star  \,\, $ & $1.6^{+1.6}_{-1.} $ & $ \phantom{x} c^2 \times  (a+a_{cb})^2$
\\[2pt]
$ W \, W^\star  \,\, $  & $0.2^{+0.6}_{-0.6} $ & $\phantom{x} c^2 \times a^2$
\\[2pt]
\hline
$ $ &$ $ &$ $
\\[-10pt]
$  \gamma \, \gamma $ & $2.1^{+2.}_{-1.6} $ & $ \phantom{x} c^2 \times r_{\gamma\gamma}(a,c)$
\\[2pt]
$  \gamma \, \gamma $ & $0.6^{+1.}_{-0.9}$ & $  \phantom{x} c^2 \times r_{\gamma\gamma}(a,c)$
\\[2pt]
$  \gamma \, \gamma $ & $2.2^{+1.5}_{-1.4} $ & $  \phantom{x} c^2 \times r_{\gamma\gamma}(a,c)$
\\[2pt]
$  \gamma \, \gamma $ & $0.6^{+1.8}_{-1.7}$ & $  \phantom{x} c^2 \times r_{\gamma\gamma}(a,c)$
\\[2pt]
$ \gamma \, \gamma  \, jj \,\, $   & $3.6^{+2.2}_{-1.6} $& $ \phantom{x} r_{\gamma\gamma jj}(a,a_{cb},c) \times r_{\gamma\gamma}(a,c)$
\\[2pt]
$  Z \, Z^\star \,\,  $  & $0.6^{+1.}_{-0.6} $  & $ \phantom{x} c^2 \times (a+a_{cb})^2$
\\[2pt]
$  W \, W^\star  \,\,  $& $0.4^{+0.6}_{-0.6} $  & $ \phantom{x} c^2 \times a^2$
\\[2pt]
$  b \, \bar{b}  \,\,  $& $1.2^{+2.1}_{-1.9}  $  & $ \phantom{x} r_{Vh}(a,a_{cb}) \times c^2$
\\[2pt]
$  \tau \, \bar{\tau}  \,\,  $& $0.6^{+1.1}_{-1.2} $  & $ \phantom{x} c^2 \times c^2$
\\[2pt]
\hline
$ $ &$ $ &$ $
\\[-10pt]
$ W \, W^\star  \,\, $  & $0.0^{+1.2}_{-0} $ & $ \phantom{x} c^2 \times a^2$
\\[2pt]
$  b \, \bar{b}  \,\,  $& $2.0^{+0.8}_{-0.7}  $  & $ \phantom{x} r_{Vh}(a,a_{cb}) \times c^2$
\end{tabular}
\vspace{10mm}
\end{minipage}
\caption{Summary table of the experimental results that we included in our analysis. The signal strengths for all CMS and Tevatron channels, as well as for the ATLAS $WW$ and $\gamma \gamma_{FP}$ are taken at $m_{h}=125\,\mathrm{GeV}$. On the other hand, for the ATLAS $ZZ$ and $\gamma \gamma$ channels we use the peak signal strength. We report the leading scaling with the parameters $(a,a_{cb},c)$ both for production cross section and partial decay width in the various channels. The predictions of the best fit points are also shown in orange.}
\label{table:tcorrections} \vspace{-0.35cm}
\end{figure}
After computing production cross sections and BRs we construct a $\chi^2$ function
\begin{equation}
  \chi^2(a,a_{cb},c)=\sum_{i} \frac{(\hat{\mu}_{i} - \mu_{i}(a,a_{cb},c))^{2}}{\delta \mu_i^2},
\end{equation}
where $\hat{\mu}_{i}$ is the experimental central value, and $\delta \mu_i$ is the total error. The latter is obtained by summing in quadrature the experimental error (symmetrized by means of an average in quadrature) to the theoretical error. The theoretical error comes from the uncertainties on cross sections, and is relevant only when two or more production mechanisms are summed over. We simply propagate the errors, taking their values for the single production mechanisms from Ref.~\cite{HiggsTwiki}.

\begin{figure}[t]
 \begin{center}
   \includegraphics[width=.45\textwidth]{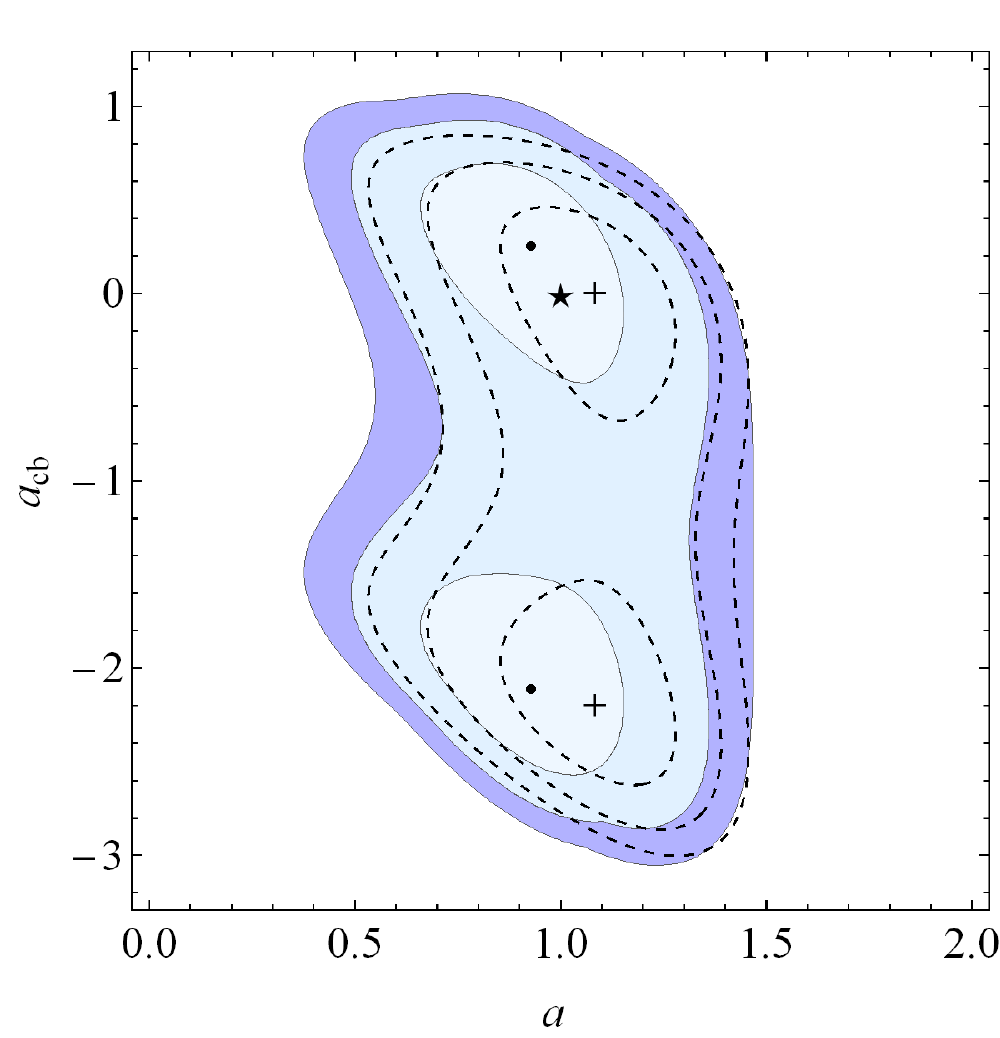}\, \, \includegraphics[width=.45\textwidth]{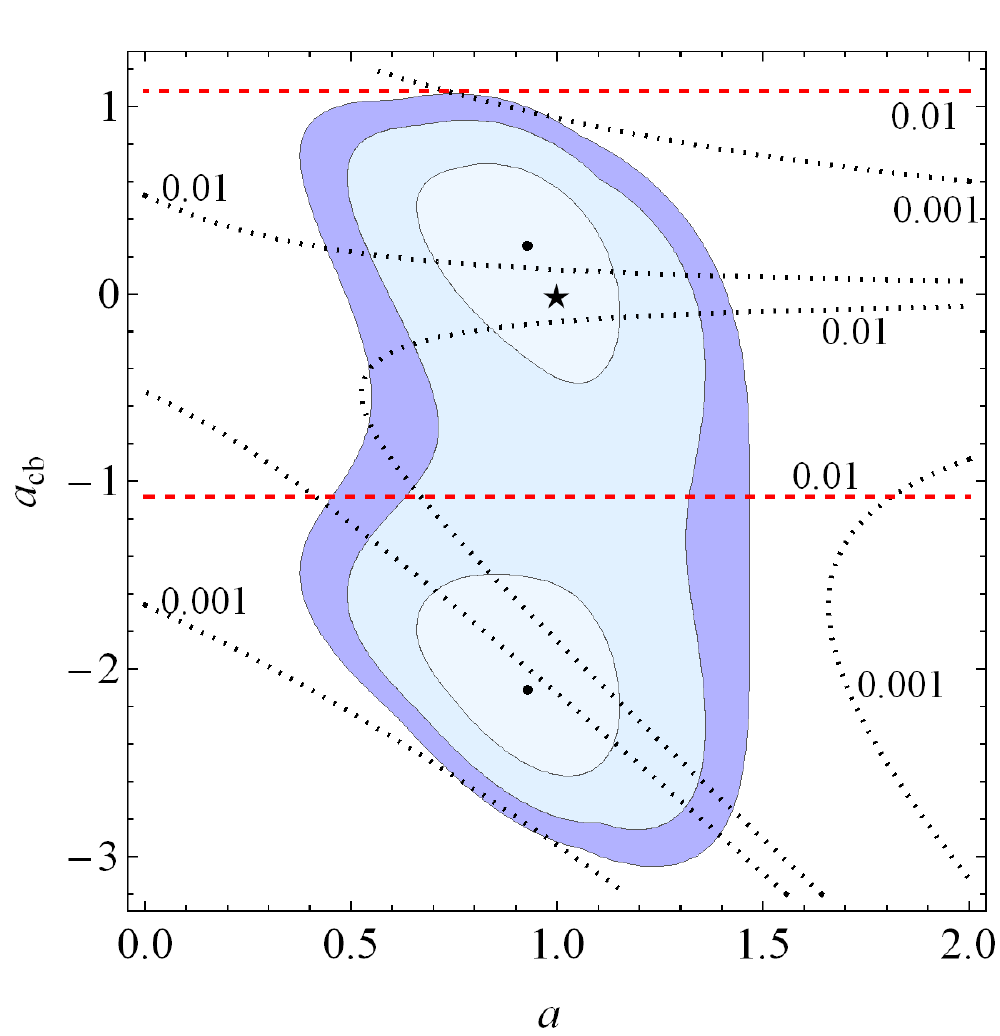}
 \end{center}
\caption{Left panel: best-fit region in the $(a,a_{cb})$ plane from LHC results, as in Fig.~3, at $68,95,99\%$ C.L. after marginalization. Dashed lines represent the analogous contours in the case $c=1$. The two best fit points with (without) marginalization are shown as black dots (crosses), while the star is the SM point corresponding to $(a,a_{cb})=(1,0)$. All the observables involved are insensitive to the sign of $a+a_{cb}$, implying the symmetry under \mbox{$(a,a_{cb})\to (a,-(2a+a_{cb}))$}. Right panel: isocontours of $|\Delta\epsilon_{1}^{UV}/\epsilon_{1}^{exp}|^{-1}\,$ (dotted, black) and of $|\Delta\epsilon_{1}^{TL}/\epsilon_{1}^{exp}|^{-1}\,$ (red, dashed), indicating the level of tuning needed to satisfy EWPT, are superimposed to the LHC best fit region.}
\label{fig:resultsLHC}
\end{figure}
Since we are interested in the gauge sector, and in particular in custodial breaking effects, we treat $c$ as a nuisance parameter. Thus a $\chi^2$ restricted to $(a,\, a_{cb})$ can be computed by marginalizing over $c\,$:
\begin{equation}
  \chi^2(a,a_{cb})= \min_{\{c\}} \chi^2(a,a_{cb},c) \, ,
\end{equation}
and it can be used to perform a minimum $\chi^2$ procedure. The result of the fit is summarized in Fig.~\ref{fig:resultsLHC} left, where we also show for completeness the results without marginalization (fixing $c=1$). The best fit points are respectively $(a,a_{cb})=(0.93, 0.25)$ and $(0.93, -2.11)$, both corresponding to $\chi^2=9.2$ with 13 d.o.f. As expected the best fit points are `$Z$philic' (or equivalently, $W$phobic): $\mu_{ZZ}/\mu_{WW}=(\cos^{2}\theta_{W}\, g_{hZZ}/g_{hWW})^{2}=(a+a_{cb})^{2}/a^{2} \approx 1.6$.

Notice that all the observables involved in Higgs searches are insensitive to the sign of $a+a_{cb}$ (as such combination always appears squared), implying the symmetry of the contours under \mbox{$(a,a_{cb})\to (a,-(2a+a_{cb}))$}. In the best-fit region where $a+a_{cb}<0\,$, the Higgs is actually `dys$Z$philic', since the sign of the $hZZ$ coupling is opposite with respect to the standard case. We will discuss in Section \ref{sec:implication} some future measurements that may lift the degeneracy between a $Z$philic and a dys$Z$philic Higgs.

As we have already mentioned in Section~\ref{Tparameter}, new light degrees of freedom are required in order to make a sizeable $a_{cb}$ compatible with EWPT. In the absence of a symmetry a significant tuning is generically needed, as shown in the right panel of Fig~\ref{fig:resultsLHC}. In principle, such new light degrees of freedom could affect the Higgs couplings, and therefore alter the interpretation of results of Higgs searches.

\begin{figure}[t]
 \begin{center}
   \includegraphics[width=.4\textwidth]{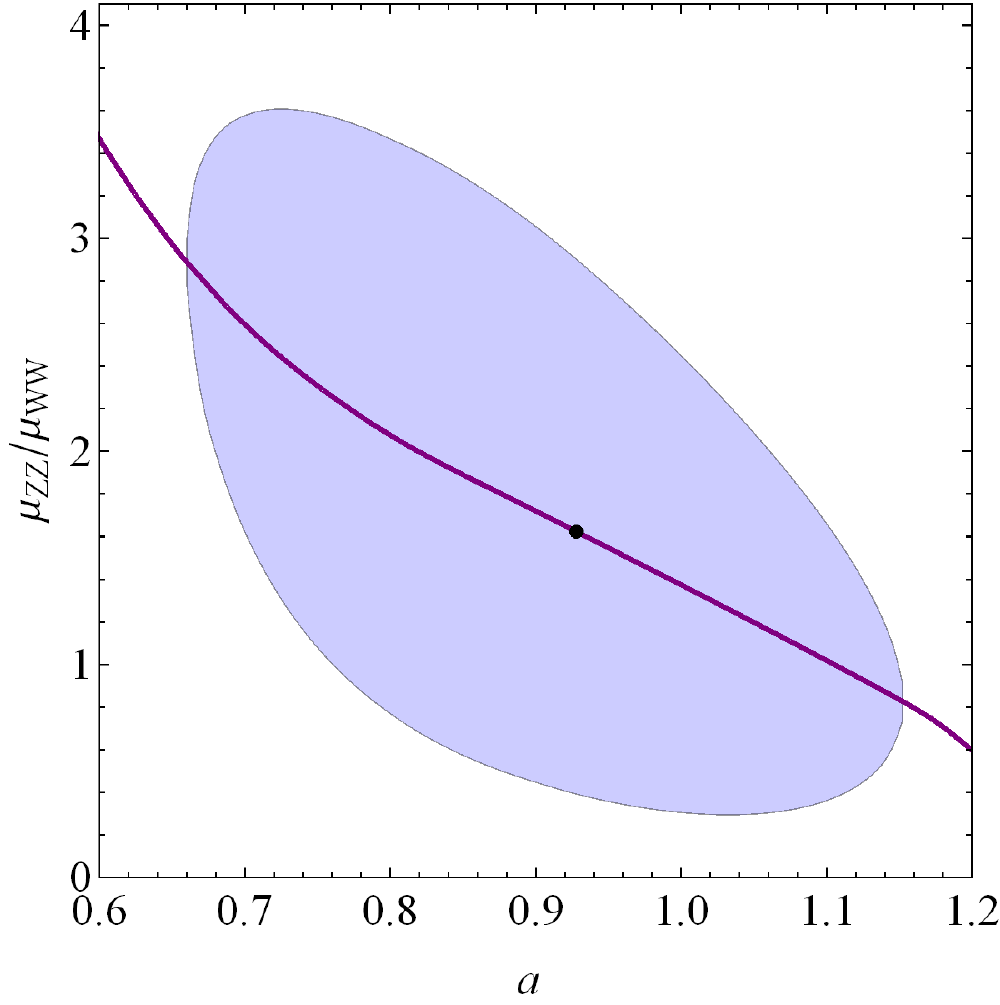} \,\,\,\,\,\,\,\, \includegraphics[width=.4\textwidth]{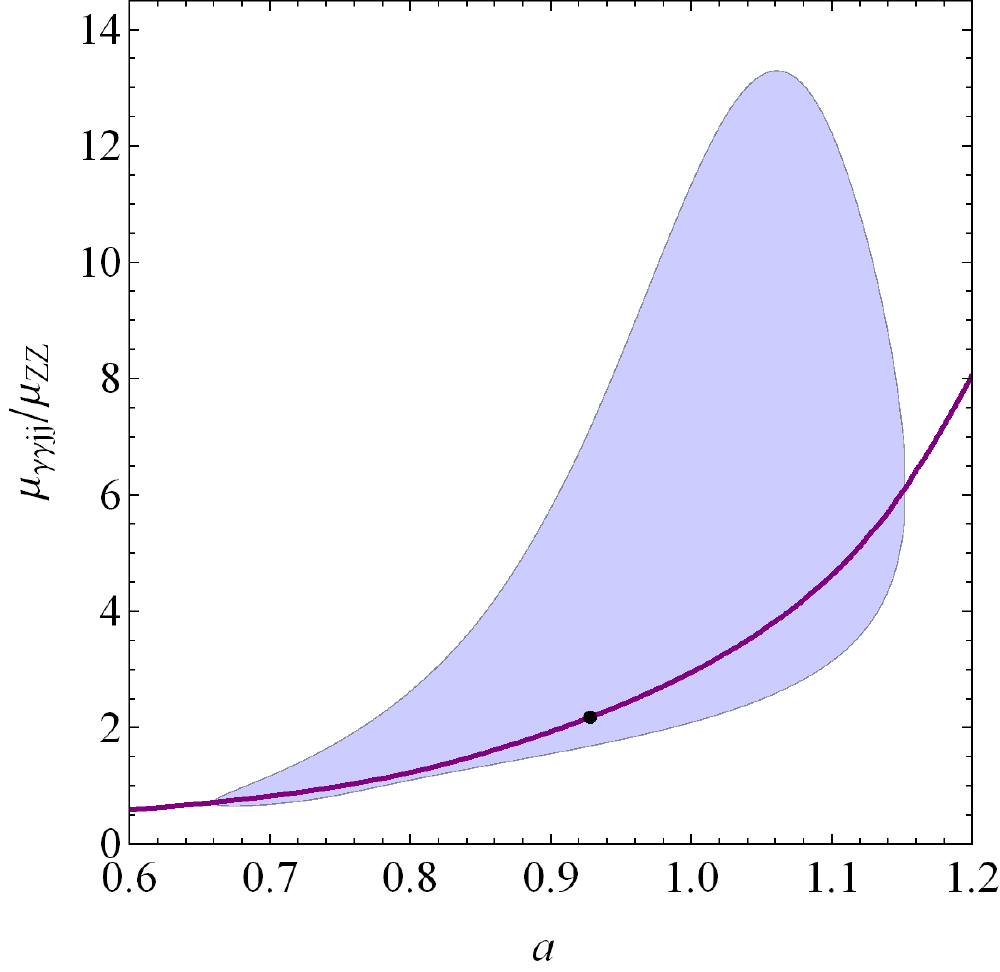}
 \end{center}
\caption{The colored regions show the range of $\mu_{ZZ}/\mu_{WW}$ (left panel) and $\mu_{\gamma\gamma jj}/\mu_{ZZ}$ (right panel) as a function of $a$, obtained varying $a_{cb}$ within the $68\%$ CL region of the LHC fit, whereas the full line corresponds to choosing the best-fit value of $a_{cb}$ for the given $a\,$.}
\label{fig:ggZZandWWZZ}
\end{figure}

An obvious consequence of $a_{cb}\neq 0$ is that the ratio $\mu_{ZZ}/\mu_{WW}$ differs from unity. This is shown in Fig.~\ref{fig:ggZZandWWZZ}, where we plot for each value of $a$ the range of $\mu_{ZZ}/\mu_{WW}$ obtained varying $a_{cb}$ within the $68\%$ CL region of the LHC fit (colored region). We see that within the LHC preferred region the wide range $0.3\lesssim \mu_{ZZ}/\mu_{WW} \lesssim 3.5$ is obtained, with the possibility of a severe $Z$philia (although $Z$phobia cannot be totally excluded at the moment).

Another channel that can be effectively enhanced is $\gamma\gamma$, due to both $c$ and $a_{cb}$. For example if $\mu_{\gamma\gamma jj}/\mu_{ZZ}$ is considered, dramatic effects are possible even within the LHC $68\%$ C.L. region, as can be seen in Fig.~\ref{fig:ggZZandWWZZ}.


\subsection{Signal strength ratios at the LHC}
We have to stress that the $(a,c)$ and $(a,a_{cb},c)$ parameterizations are different and in principle it is possible to distinguish between them. The best way is to look at ratios between well measured $\mu_i$, as most of the QCD production uncertainties are thus cancelled (especially if the production channel is the same), as well as the dependence on the total width. See Refs.~\cite{Espinosa:2012ir,Azatov:2012rd} for a discussion of how to break degeneracies in similar fits by using ratios of signal strengths.

\begin{figure}[t]
 \begin{center}
      \includegraphics[width=.45\textwidth]{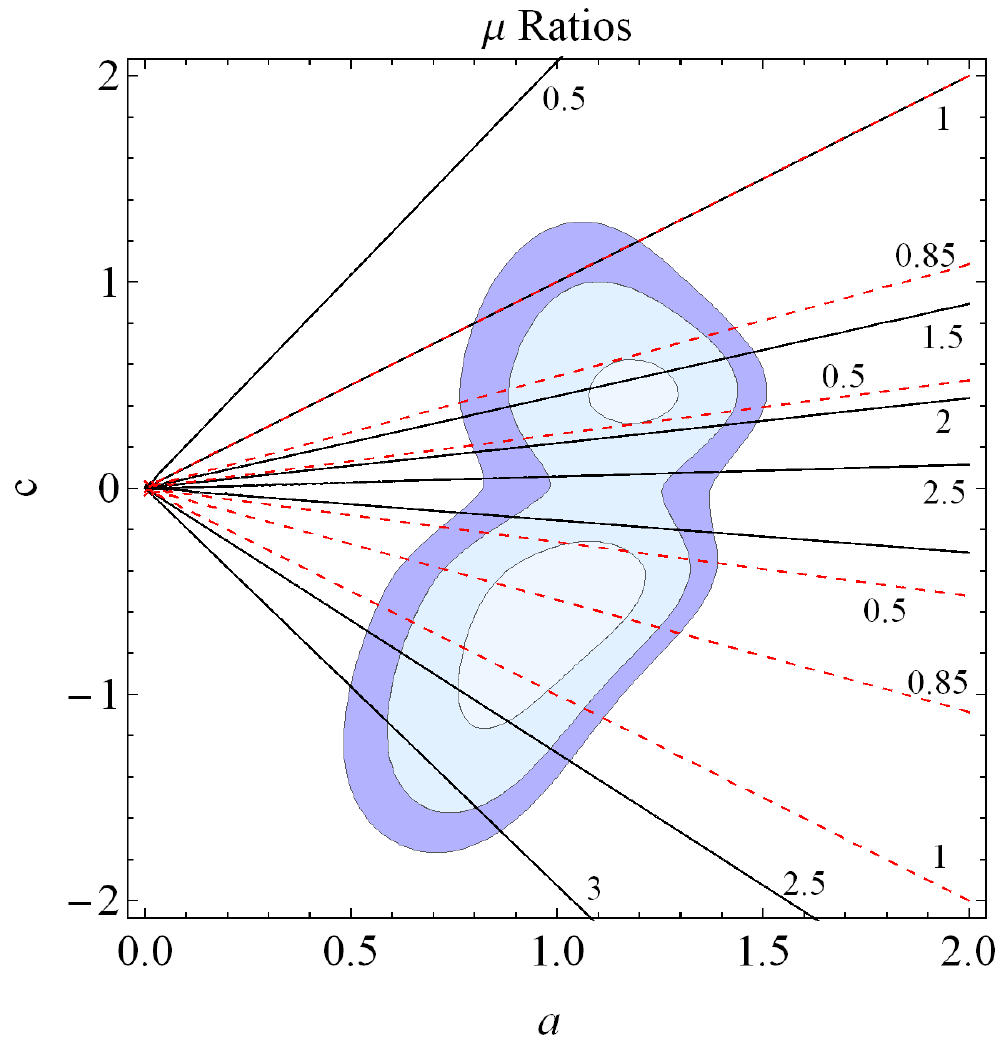}   \includegraphics[width=.45\textwidth]{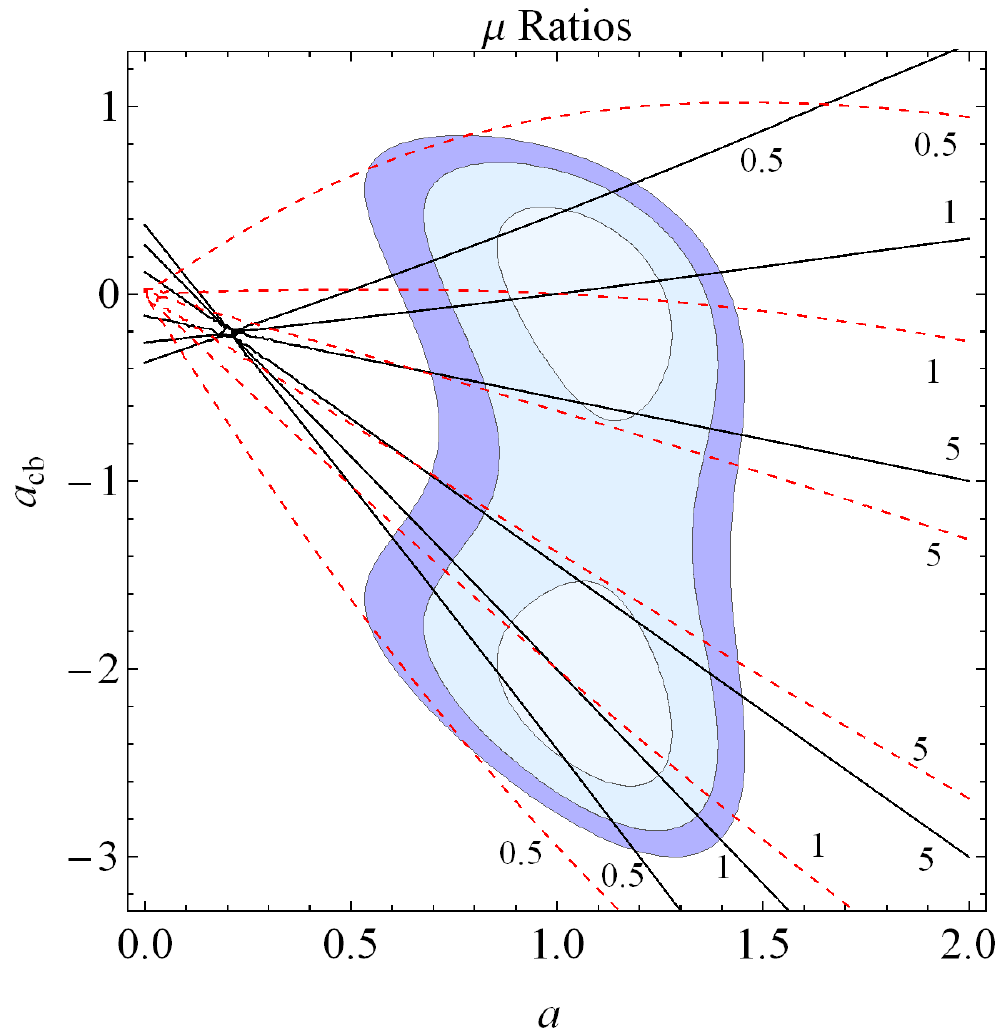}
 \end{center}
\caption{Isocurves of $\mu_{\gamma \gamma}/\mu_{ZZ}$ (solid) and of $\mu_{b \bar{b}}/\mu_{ZZ}$ (dashed) in the $(a,c)$ plane (left panel) and in the $(a,a_{cb})$ plane (right panel). In both plots the LHC best-fit regions are also shown; in the right panel, $c=1$ has been set to facilitate the comparison with the custodial-preserving case.}
\label{fig:acVSaacb}
\end{figure}

To show how it can be possible to distinguish between the different cases, we choose the ratios \mbox{$(\mu_{\gamma \gamma}/\mu_{ZZ},\mu_{b \bar{b}}/\mu_{ZZ})$}: in Fig.~\ref{fig:acVSaacb} we show isocurves of such ratios in the $(a,c)$ and $(a,a_{cb})$ planes respectively, superimposing them to the LHC best fit regions. To simplify the comparison, in the right panel of Fig.~\ref{fig:acVSaacb} we have set $c=1$. We see that in the $(a,c)$ case the range allowed for the ratios is significantly smaller than it is in the  custodial breaking case.

\section{Future implications}\label{sec:implication}
We are left with the issue of determining the sign of $a_{cb}$ (or of $a+a_{cb}$ if you prefer). Not an easy quest, as the sign is physically relevant only in the presence of interference. One readily available choice would be to look again at precision tests, in this case corrections to the $Zb\bar{b}$ vertex. However the ratio between the main 1-loop Higgs contributions and the one of interest for us goes as $m_t^2/m_b^2$ and so we expect the latter to be negligible. Thus we have to turn our attention to other, not yet measured, processes. We are going to briefly discuss four possible experimental signatures that are, or can be in principle, sensitive to the sign of the $hZZ$ coupling.
\par
Before moving to a discussion of the single channels, few comments are in order. We are interested in processes where diagrams both with and without the $hZZ$ vertex interfere, and we need such interference to be non negligible in order to distinguish between the two signs. Let us stress that the separation has to be bigger than both experimental and theoretical uncertainties. Concerning the latter, a precise knowledge of the absolute value of the coupling constants ($a$ and $a+a_{cb}$ in particular) is required. Thus we are going to focus on possible scenarios at $e^+ e^-$ Linear Colliders (LC), for which it is reasonable to assume a measurement of $g_{hZZ}$ and $g_{hWW}$ at the level of $\sim 1\%$, corresponding to
\beq
|\delta a|, \, |\delta(a+a_{cb})| \sim 1 \% \, .
\eeq
Such precision is expected both at ILC \cite{Djouadi:2007ik}  and CLIC \cite{Linssen:2012hp} with reference values \mbox{$m_h=120\,\mathrm{GeV}$}, $\sqrt{s}=500$ GeV and with $500$ $\mathrm{fb}^{-1}$ of integrated luminosity. In the following we fix $c=1$ in order to highlight the main points under study.

\subsection{$h \rightarrow ZZ$ decay width}
\begin{figure}[t]
 \begin{center}
   \includegraphics[width=.75\textwidth]{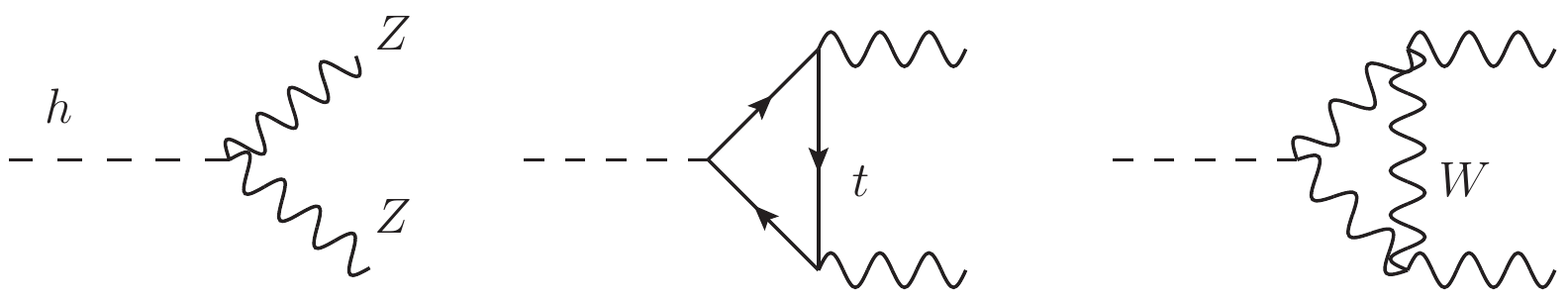}
 \end{center}
\caption{Leading order and main NLO contributions to $h \to ZZ$.}
\label{h to ZZ diagrams}
\end{figure}

The first channel we investigate is the width of $h \rightarrow ZZ \to 4l$. Here the interference occurs between tree level and higher orders, the former being sensitive to the sign flip \mbox{$a+a_{cb}\to -(a+a_{cb})$}. On the contrary we assume, in order to maximize the separation, that most of the radiative corrections arise from loops not directly involving the $hZZ$ vertex (see the diagrams in Fig.~\ref{h to ZZ diagrams}). In this approximation the two cases $a+a_{cb} \gtrless 0$ have different relative sign between LO and NLO. Thus we can write the width in the two cases as (the superscript corresponds to the sign of $a+a_{cb}$) $\Gamma_{ZZ}^\pm \approx \Gamma_{ZZ}^0(1\pm \delta)$, with $\delta\approx 1 \%$ for SM couplings \cite{Bredenstein:2006rh}.
Assuming departures from the leading approximation $a+a_{cb} =\pm 1\,$ to have negligible effects, we quantify the relative separation with
\beq
\Delta=\left|\frac{\Gamma_{Z}^{+}-\Gamma_{Z}^{-}}{\Gamma_{Z}^{+}+\Gamma_{Z}^{-}}\right|=  \delta \approx 1 \% \, .
\eeq
It is clear that a very high precision is required to resolve the two cases. In fact, even considering perfect knowledge of the coupling constants, the experimental uncertainties should be at least of the same size or smaller of $\Delta$. We conclude that the measurement under study is not realistic.

\subsection{$ht \bar{t}$ associated production}
\begin{figure}[t]
 \begin{center}
   \includegraphics[width=.75\textwidth]{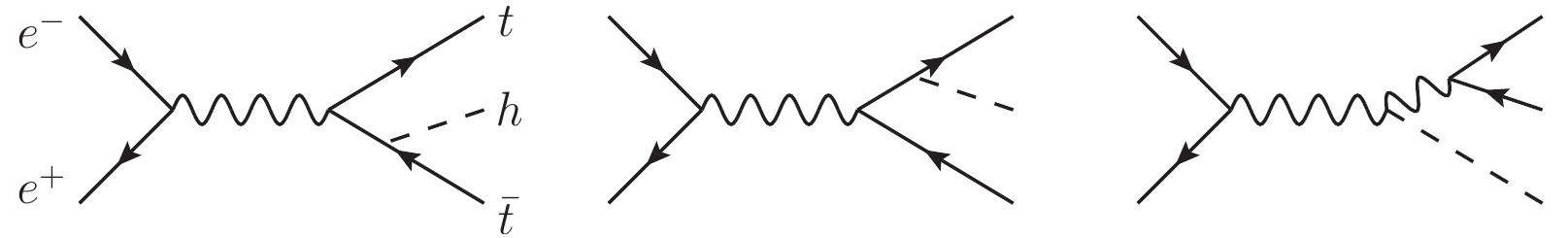}
 \end{center}
\caption{Feynman diagrams contributing to $e^{+}e^{-}\to ht\bar{t}$.}
\label{tth diagrams}
\end{figure}

We now focus on a case where the interference arises between different LO contributions. In Higgs boson associated production with tops (heavy fermions in general) the process is essentially $e^+ e^- \rightarrow Z \rightarrow t \bar{t}$ with a scalar emitted either by the $Z$ or by one of the tops (as shown in Fig.~\ref{tth diagrams}).
We can write the total cross section for the two cases $a+a_{cb}=\pm 1$ as follows
\beq
\sigma_{\pm} =(\sigma_{t}+\sigma_{Z} \pm \sigma_{int})\,,
\eeq
where the index refers to the particle the Higgs boson is emitted from. We have $\sigma_{int}/(\sigma_{t}+\sigma_{Z}) \approx 1-4\%\,$, leading to
\beq
\Delta= \left|\frac{\sigma_+-\sigma_-}{\sigma_++\sigma_-}\right| \lesssim 4 \% \, ,
\eeq
that needs to be compared to the experimental resolution. It has been shown \cite{Gay:2006vs} that from  $e^+ e^- \rightarrow t \bar{t}$ the coupling $g_{tth}$ could be measured up to $6\%$ precision\footnote{At ILC with $\sqrt{s}=800$ GeV and with $1000$ $\mathrm{fb}^{-1}$ of integrated luminosity.}, which directly translates in a precision of around $10-12\%$ on the cross section, at least 3 or 4 times larger than $\Delta$. So even this case seems unlikely to be able to resolve the different signs.

\begin{figure}[t]
 \begin{center}
 \includegraphics[width=.75\textwidth]{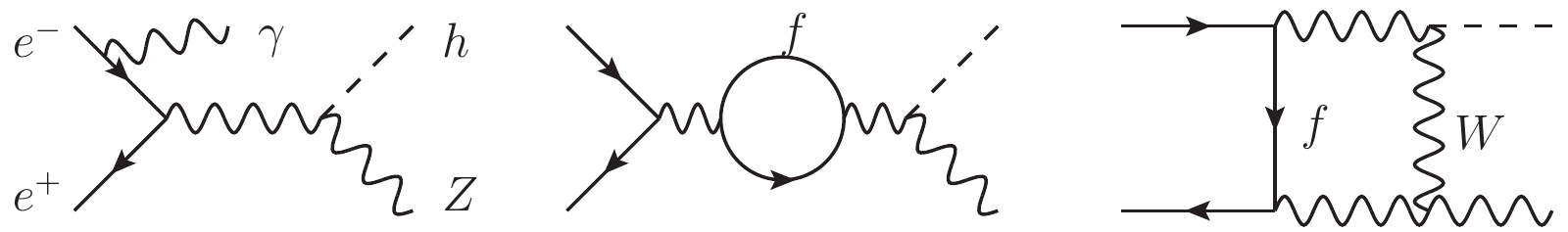}
 \end{center}
\caption{Representative diagrams for each of the 3 classes of radiative corrections to \mbox{$e^{+}e^{-}\to Zh$}, see text for details.}
\label{Zh diagrams}
\end{figure}
%
%

\subsection{$Z h$ associated production}

\begin{figure}[t]
 \begin{center}
   \includegraphics[width=.75\textwidth]{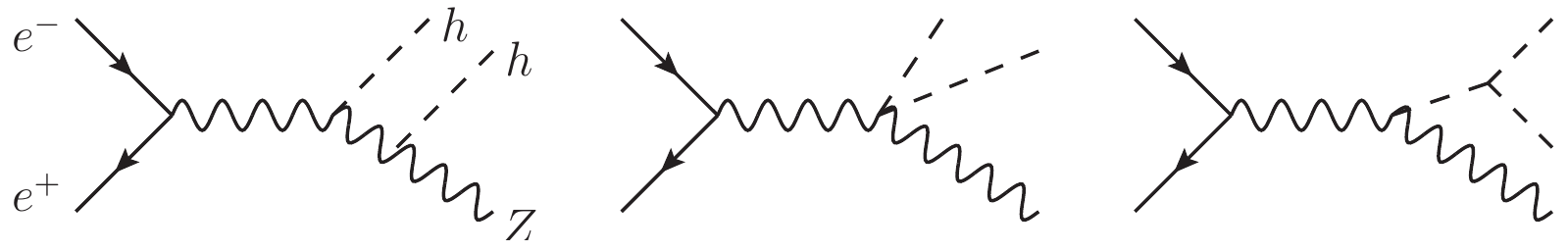}
 \end{center}
\caption{Feynman diagrams contributing to $e^{+}e^{-}\to Zhh$.}
\label{Zhh diagrams}
\end{figure}

The third channel we examine is the Higgs-strahlung process $ e^+ e^- \to Z h$, see Fig.~\ref{Zh diagrams}. As in the first case above, we are interested in the change in sign of NLO corrections with respect to the tree level amplitude. Following detailed analyses present in literature \cite{Kniehl:1991hk,Denner:1992bc} we can divide the main electroweak corrections in three different terms, as following:
\begin{itemize}
  \item Initial State Radiation ($\delta_{ISR}$): whose amplitude clearly has the same sign of the tree level one;
  \item Fermionic contributions ($\delta_F$): they are mainly due to self energy corrections to the Z propagator. Thus, in first approximation, we expect them to have the same sign of the LO amplitude;
  \item Bosonic contributions ($\delta_B$): they are due to box diagrams usually involving $W$ bosons. It is reasonable to assume that most of these would not involve the $hZZ$ vertex and so to assume that $\delta_B$ does not present a sign flip for a dis$Z$philic Higgs.
\end{itemize}
It is then possible to write, for $a+a_{cb} \gtrless 0$,
\beq
\sigma_\pm=\sigma_0(1+\delta_{ISR}+\delta_F\pm\delta_B) \,
\eeq
as a rough estimate of the effect. Referring to a center of mass energy of $1\,\mathrm{TeV}$ the expected magnitudes for such corrections\footnote{The numerical values are extracted from Ref.~\cite{Denner:2003yg}, where $m_{h}= 150$ GeV was assumed. However, corrections due to the lower Higgs mass we are considering should be small and nonetheless would not change our conclusions.} are $\delta_{ISR} \approx 20\%$, $\delta_F \approx 10\%$ and $\delta_B \approx - 20\%$. Thus $\sigma_+ \approx 1.1 \sigma_0$, $\sigma_- \approx 1.5 \sigma_0$ and we are able to quantify the separation between the two cases as
\beq
\Delta = \left|\frac{\sigma_+-\sigma_-}{\sigma_++\sigma_-}\right| \approx 15 \% \, ,
\eeq
if we consider the simple choices $a+a_{cb}=\pm 1\,$. A comparison with the expected experimental sensitivity \cite{GarciaAbia:1999kv}, which is of $\sim 3-5\%$, shows that this measurement would indeed be able to resolve the sign.
\subsection{$Zhh$  production}
Another process where interference is at leading order is $e^+ e^- \rightarrow Z \rightarrow Zhh\,$. In this case there are three distinct constributions: the diagram with two subsequent Higgs-strahlungs, the diagram involving the $hhZZ$ vertex, and a third one involving the Higgs self-coupling (see Fig.~\ref{Zhh diagrams}), the last being the only one that changes sign under $(a+a_{cb})\to - (a+a_{cb})\,$. The cross section for $a+a_{cb}=\pm 1$ can then be written as
\beq
\sigma_{\pm} =\sigma_{0} \pm \sigma_{int}\,,
\eeq
and for $\sqrt{s}=500\,\mathrm{GeV}$ (which is the best choice for the process $e^{+}e^{-}\to Zhh$) we find \mbox{$\sigma_{+}\simeq 0.28\, \mathrm{fb}$}, \mbox{$\sigma_{-}\simeq 0.09\,\mathrm{fb}$}. Therefore
\beq \label{delta Zhh}
\Delta= \left|\frac{\sigma_{+} - \sigma_{-}}{\sigma_{+} +\sigma_{-}}\right| \approx 50 \% \, ,
\eeq
that needs to be compared to the experimental resolution. For an integrated luminosity of $2000\,\mathrm{fb}^{-1}$ and SM couplings, this can be as low as $10\%$ \cite{CastanierETAL}. In the case of flipped $hZZ$ coupling, by taking into account the reduced statistics we estimate the resolution to be still less than $20\%$, i.e. more than two times smaller than $\Delta$. So this case is promising. However, we warn the reader that in the previous discussion we have made stronger assumptions than for the other precision measurements we presented. First, when setting the Higgs self-coupling $\lambda_{hhh}$ to its SM value, we assumed to know it to a good accuracy, even though the measurement of such coupling at the LHC would be a difficult task, and the best channel to measure the trilinear at a LC with moderate $\sqrt{s}\,$ would be $e^{+}e^{-}\to Zhh$ itself (an independent measurement of $\lambda_{hhh}$ could come from the $WW$ fusion process $e^{+}e^{-}\to \nu\bar{\nu}hh\,$ at $\sqrt{s}\sim 1\,\mathrm{TeV}\,$). Second, we assumed the $hhZZ$ coupling to have its SM value although its measurement is challenging even at a LC, and despite the fact that in a theory with $(a,a_{cb})\neq (1,0)$ we should in general expect deviations from the standard values also in the couplings $hhZZ$ and $hhWW$. As a consequence, one should take the estimate in Eq.~\eqref{delta Zhh} with some caution.

\section{Conclusions}
Motivated by recent results of experimental searches for the Higgs boson, we relaxed the assumption of custodial invariance in its couplings to the $W$ and the $Z$. We described custodial breaking through an additional parameter $a_{cb}$ and we showed how it can accommodate the current pattern of observed excesses, which mildly point to a $Z$philic (or $W$phobic) Higgs. Should such hints be confirmed by more data, they would be evidence for custodial breaking in Higgs couplings. Such breaking implies that the electroweak $T$ parameter receives quadratically divergent corrections. New light degrees of freedom would then be expected to play a role in mimicking the approximate custodial invariance observed in EW data, generically at the price of a sizable tuning.

We also noticed that Higgs searches are insensitive to the sign of the $hZZ$ coupling, that is to say they do not allow to tell a $Z$philic Higgs from its dys$Z$philic counterpart. However the sign of such coupling is physical, and processes in which interference is present can remove the degeneracy. We presented some measurements at future linear colliders that could be used for this purpose.

\section*{Acknowledgments}
We are grateful to A.~Falkowski and G.~Isidori for discussions about the presence of quadratic divergences in $T$ in the absence of custodial symmetry. We thank \mbox{R.~Contino} and M.~Trott for comments about the manuscript, and \mbox{J.~R. Espinosa}, A.~Juste and D.~Pappadopulo for discussions. We also thank the authors of Ref.~\cite{Giardino:2012ww} for inspiring the title of this work. This research has been partly supported by the European Commission under the ERC Advanced Grant 226371 {\it MassTeV} and the contract PITN-GA-2009-237920 {\it UNILHC}. E.~S. has been supported in part by the European Commission under the ERC Advanced Grant 267985 {\it DaMeSyFla}. The work of  M.~F. has been partly supported by the \emph{Fondazione A.~Della Riccia}.

\end{document}